**Interlayer Topological Transport and Devices based on Layer Pseudospins in Photonic Valley-Hall Phases**

*Xiaoxiao Wu, Zhenyu Li, Jian Chen, Xin Li, Jingxuan Tian, Yingzhou Huang, Shuxia Wang, Weixin Lu, Bo Hou[*], C. T. Chan, Weijia Wen[*]*

Dr. X. Wu, Prof. C. T. Chan, Prof. W. Wen
Department of Physics,
The Hong Kong University of Science and Technology,
Clear Water Bay, Kowloon, Hong Kong, China
E-mail: phwen@ust.hk (Prof. W. Wen)

Z. Li, J. Chen, Prof. W. Lu, Prof. B. Hou
School of Physical Science and Technology & Collaborative Innovation Center of
Suzhou Nano Science and Technology,
Soochow University,
Suzhou 215006, China
E-mail: houbo@suda.edu.cn (Prof. B. Hou)

X. Li, Prof. Y. Huang, Prof. S. Wang
Chongqing Key Laboratory of Soft Condensed Matter Physics and Smart Materials,
College of Physics,
Chongqing University,
Chongqing, 400044, China

J. Tian
Department of Mechanical Engineering, Faculty of Engineering,
The University of Hong Kong,
Hong Kong, China

Prof. W. Lu, Prof. B. Hou
Key Laboratory of Modern Optical Technologies of Ministry of Education & Key Lab
of Advanced Optical Manufacturing Technologies of Jiangsu Province,
Suzhou 215006, China

Prof. W. Wen
Materials Genome Institute,
Shanghai University,
Shanghai 200444, China

**Abstract**

Valley-Hall phases, first proposed in two-dimensional (2D) materials, originate from nontrivial topologies around valleys which denote local extrema in momentum space. Since they are extended into classical systems, their designs draw inspirations from existing quantum counterparts, and their transports show similar topological protections. In contrast, it is recently established in acoustics that layer pseudospins in valley-Hall phases can give rise to special valley-Hall edge states with fundamentally different transport behaviors at the interfaces compared with various 2D materials. Their realization in other classical systems, such as photonics, would allow us to design topological insulators beyond quantum inspirations. In this work, we show that layer pseudospins exist in photonic valley-Hall phases, using vertically coupled designer surface plasmon crystals, a non-radiative system in open environment supporting tightly-confined propagating modes. The negligible thermal and radiative losses in our structure pave the way for our direct observations of the layer pseudospins and associated topological phenomena stemmed from them in both real and reciprocal spaces. Photonic devices that manipulate signals based on layer pseudospins of the topological phases, such as layer convertors and layer-selected delay lines, are experimentally demonstrated, confirming the potential applications of layer pseudospins as a new degree of freedom carrying information.

## 1. Introduction

In the last decade, extensive pursuits of valley-dependent transports in two-dimensional (2D) materials — valleys denoting local extrema in momentum space[1, 2] — have led to a new type of electronics termed 'valleytronics'[3, 4]. It focuses on bulk-gapped valley-Hall phases whose nontrivial topology around well-separated valleys is the key to generate valley-polarized currents[5-7]. Recently, valley-Hall phases, like previous Chern[8-12], Floquet[13-17] and spin-Hall phases[18-23], have been extended to classical bosonic systems. Classical valley-polarized edge states as the analogy of valley-polarized currents, including their topologically protected transports, have been successfully observed[24-29]. Devices based on classical valley-Hall phases are also demonstrated[30-32]. Remarkably, layer pseudospins in acoustic valley-Hall phases can give rise to topological transport behaviors currently unfound in 2D materials[33], and they may become a new degree of freedom (DOF) carrying information for their easy detections by external probes. Do layer pseudospins and their extraordinary transports exist in other bosonic systems? If so, we could start designing classical topological systems beyond inspiration from quantum systems

In this Article, we report a scheme to realize layer pseudospins in photonic valley-Hall phases using vertically coupled designer surface plasmon (DSP) crystals, where tightly-confined modes propagate on periodically patterned metallic surfaces[25, [34-40]. Valley-Hall phases and layer pseudospins are generated from twisting of anisotropic slots. Through near-field mappings, both layer-mixed and layer-polarized valley-Hall edge states characterized by a pair of half-integer quantized valley Chern numbers (VCNs) are directly observed and resolved. Unique devices which manipulate photonic valley-Hall edge states based on layer pseudospins are demonstrated.

## 2. Results and Discussion

Our system is a triangular-lattice crystal shown in **Figure 1**(a), and the inset shows its first Brillouin zone (FBZ). A unit cell outlined by the red hexagon is detailedly shown in Figure 1(b). This composite DSP crystal comprises three metallic layers and two dielectric spacers. A series of top-view schematics depicting the metallic layers are shown in Figure 1(c). Identical but twisted "ceiling fan"-shaped slots are arranged in triangular lattices on upper and lower metallic layers, while circular slots are arranged in a honeycomb lattice on the middle one. The orientations of "ceiling fan"-shaped slots are parameterized by relative rotation angle $\alpha$ and common rotation angle $\beta$. As a specific example, we consider following geometric parameters: lattice constant $a$ = 12 mm, metallic layer thickness $t_{metal}$ = 35 μm, spacer thickness $t_{sub}$ = 1.5 mm, and for "ceiling fan"-shaped slots, radii $R_0$ = 1 mm, $R_1$ = 5.6 mm (three longer fans), $R_2$ = 4.6 mm (three shorter fans), and for circular slots, radius $R_c$ = 2.5 mm. To reduce loss, we use F4B laminates as spacers[25, 41, 42].

We numerically calculate band structures using COMSOL Multiphysics (see Note 5 for detailed setup of the simulations, Supporting Information). The band structure when $(\alpha, \beta) = (0, 0)$ is shown in Figure 1(d), where dashed lines mark the light cone, and only bound states under it are shown. Two separated Dirac cones form around K point with a frequency split due to interlayer hoppings mediated by circular slots on the middle metallic layer. They intersect each other at an equal-frequency circle in reciprocal space, leading to a nodal-ring degeneracy (see Fig. S1 and S2 for details and 3D band structures, Supporting Information).

Generally, to induce topological effects, an essential step is to lift the degeneracies, as in our case, the nodal ring, and open a full bandgap[43]. Since nodal rings are usually protected by mirror symmetries[44], we tune $\alpha$ and $\beta$ to twist the "ceiling fan"-shaped slots, reducing mirror symmetries which open a full bandgap between the 2nd and 3rd bands (Figure 1(e)). We first only increase $\alpha$, which breaks the mirror symmetry with respect to the middle layer. It is seen the nodal ring is fully gapped once $\alpha$ becomes non-zero but the degeneracy at K point persists, protected by the two-fold rotational symmetry along ΓK direction. This point degeneracy is lifted if $\beta$ also becomes non-zero (not shown in Figure 1(e)). Alternatively, we solely increase $\beta$, which breaks the mirror symmetry with respect to ΓK direction. Initially, both Dirac cones are gapped but the nodal ring persists. Next, after $\beta$ exceeds a threshold value, the 2nd and 3rd bands detach each other, opening a full bandgap.

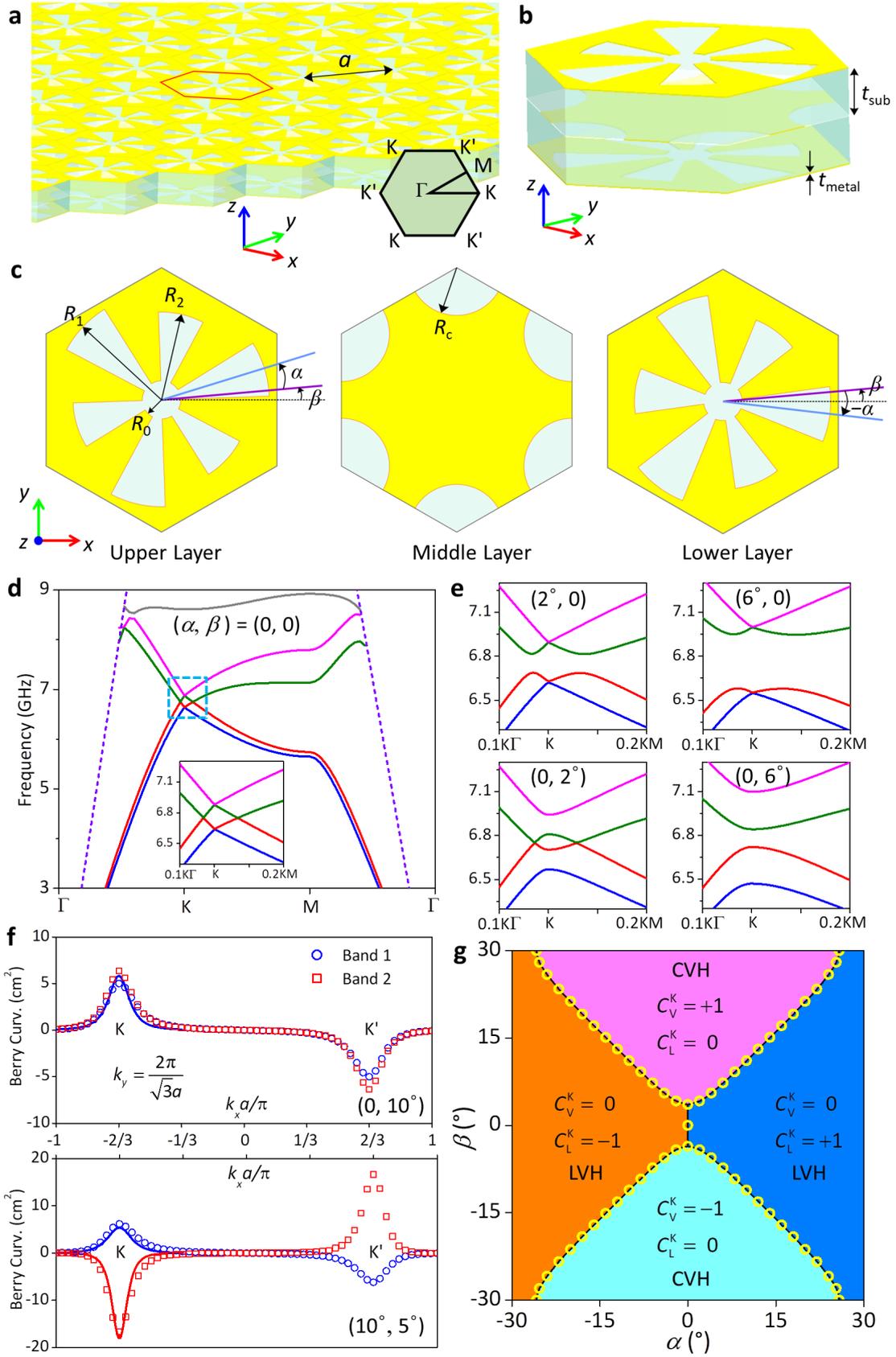

**Figure 1**. (a) Isometric schematic of the triangular-lattice composite DSP crystal, inset

showing its FBZ. The red hexagon outlines a unit cell, detailedly depicted in (b). (c) Top-view schematics of metallic layers. "Ceiling fan"-shaped slots on upper (lower) layer are rotated through the angle $\beta+\alpha$ ($\beta-\alpha$). (d) When $(\alpha, \beta) = (0, 0)$, two Dirac cones emerge around K valley, their frequency split depending on interlayer hoppings. (e) Detailed band structures around K point once $(\alpha, \beta)$ deviates from (0, 0). (f) Berry curvatures along $k_x$ when $k_y = 2\pi/(\sqrt{3}a)$. Scatters (sold lines) are calculated from COMSOL Multiphysics (perturbation Hamiltonian). (g) Phase diagram of the composite DSP crystal classified by VCNs $C_V^K$ and $C_L^K$. Yellow circles represent numerically obtained phase boundaries, where the bandgap closes. "LVH" represents layer-polarized valley-Hall phases, where fields of the modes around valleys are largely concentrated on either the upper or lower layer. "CVH" represents conventional valley-Hall phases, where fields of the modes around valleys are distributed on both layers in a relatively even manner.

We have obtained full bandgaps by tuning $\alpha$ or $\beta$ or both. However, do they belong to the same topological phase? Numerically, we calculate Berry curvatures [25, 45] of the bands below the full bandgap for two typical cases: $(\alpha, \beta) = (10°, 5°)$ and $(0, 10°)$, and the results along the line $k_y = 2\pi/(\sqrt{3}a)$ are plotted as scatters in Figure 1(f). It is seen that Berry curvatures of both cases are localized (see also 2D maps in Figure S8, Supporting Information) and largely identical for the 1st bands, but have opposite signs for the 2nd bands, which implies the two cases are possibly different topological phases. Since topological phase transitions usually happen after the close and reopen of full bandgaps[24, 25], we sweep the ($\alpha$, $\beta$) parametric space numerically and record the values when the 2nd and 3rd bands touch each other, which delineate the parametric space into four different regions (Figure 1(g)).

Now, we use **k·p** method to develop a perturbation Hamiltonian around K valley to explain our numerical findings. The perturbation Hamiltonian is spanned by the four degenerate states in the absence of twisting (intralayer) and interlayer couplings. To the first-order, it is written as [33] (see Note 1 and 2 for derivation and validation, Supporting Information)

$$\delta H_{\mathrm{K}} = v_{\mathrm{D}} s_0 \otimes (\delta k_x \sigma_x + \delta k_y \sigma_y) + \eta(\alpha s_z + \beta s_0) \otimes \sigma_z - \Delta_{\mathrm{c}} s_x \otimes \sigma_0, \qquad (1)$$

where $s_i$ and $\sigma_i$ are Pauli matrices ($i = 0$, $x$, $y$, $z$) acting on layer pseudospins and orbitals, respectively. The Hamiltonian includes three parameters, $v_{\mathrm{D}}$ group velocity, $\eta$ intralayer coupling coefficient, and $\Delta_{\mathrm{c}}$ interlayer coupling coefficient. They depend on geometry and material of the unit cell and can be extracted from numerical calculations. Solving the perturbation Hamiltonian, we also obtain Berry curvatures of the 1st and 2nd bands (solid lines in Figure 1(f)). It is observed that the first-order perturbation Hamiltonian nicely captures the essence of exact distributions of Berry curvatures, reassuring us to use the Hamiltonian to classify the band topology. Inspired by the topological invariants characterizing quantum spin Hall effect [46], we introduce a pair of VCNs [33]: besides the conventional VCN $C_{\mathrm{V}}^{\mathrm{K}}$, a layer VCN $C_{\mathrm{L}}^{\mathrm{K}}$ is also introduced, which is the difference of integrals of layer-projected Berry curvatures in one half of the FBZ (see Note 3 for formal definitions, Supporting Information). From the Hamiltonian, the half-integer quantized $C_V^K$ and $C_L^K$ are obtained and summarized in Figure 1(g). For reasons discussed below, phases with non-zero $C_L^K$ are termed layer-polarized valley-Hall (LVH) phases, and phases with non-zero $C_V^K$ are termed conventional valley-Hall (CVH) phases. From the phase diagram, we can identify $(0, 10°)$ as a CVH phase, $(10°, 5°)$ as an LVH phase, they indeed topologically different.

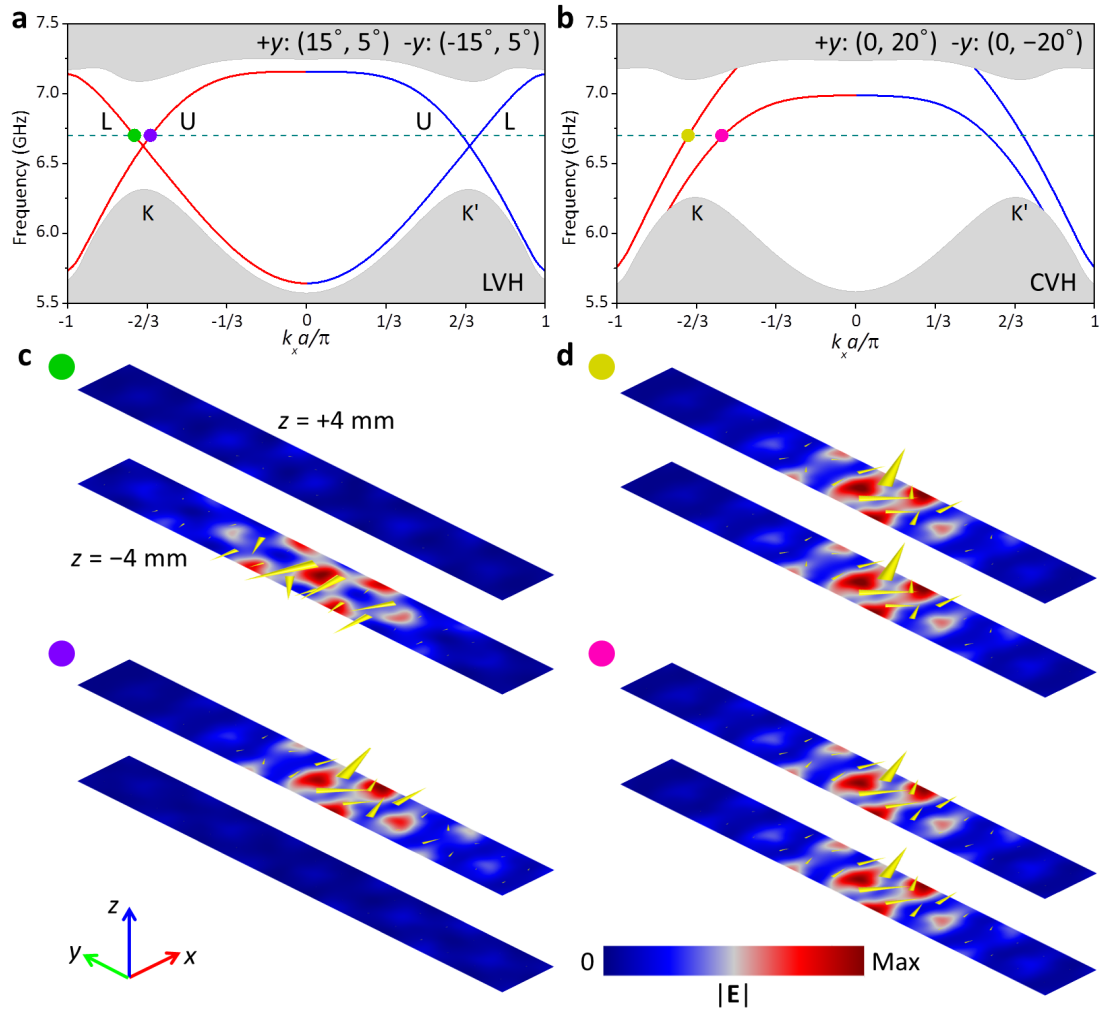

**Figure 2**. (a,b) Projected band structures for supercells comprising different LVH phases $C_L^K = \pm 1$ (a) and different CVH phases $C_V^K = \pm 1$ (b). The gray shaded regions

correspond to bulk bands. Red (blue) solid lines represent valley-Hall edge states associated with K (K') valley. Colored dots denote specific states around K valley at 6.70 GHz indicated by dashed lines. (c), (d) Simulated field maps ($|\mathbf{E}|$) of denoted states on the slices z = ±4 mm. The yellow arrows denote simulated energy flows. Edge states between LVH phases are layer-polarized, either upper (U) or lower (L), but unpolarized between CVH phases. States around K' valley can be obtained through time-reversal symmetry.

To explore possible valley-Hall edge states, two different LVH phases (±15°, 5°) are assembled to form a supercell, and its projected band structure (**Figure 2**(a)) shows two branches of valley-Hall edge states around K valley with opposite propagating directions (termed 'LVH edge states'). For comparison, we also assemble a supercell comprising two different CVH phases (0, ±20°), and its projected band structure (Figure 2(b)) shows two branches with the same propagating direction around K valley (termed 'CVH edge states'). To investigate their layer polarizations, we plot field maps of the edge states on slices $z = \pm4$ mm, namely, 2 mm above (below) the upper (lower) metallic layer, in Figure 2(c) and 2(d). Energy flows are also plotted as the arrows. As shown, LVH edge states are layer-polarized and the energy flows further indicate their layer polarizations are locked to their propagating directions around K valley. This property may be termed 'layer-chirality'. Since time-reversal transformation does not affect field distributions but reverses energy flows, the layer-chirality will be reversed at K' valley. On the contrary, there is no such property for CVH edge states, since the field distribution is not polarized to a specific layer and no layer polarization can be defined. Nevertheless, backscattering of these edge states is forbidden if no inter-valley scattering is introduced, which depends on the geometry of interfaces[24, 25, 47].

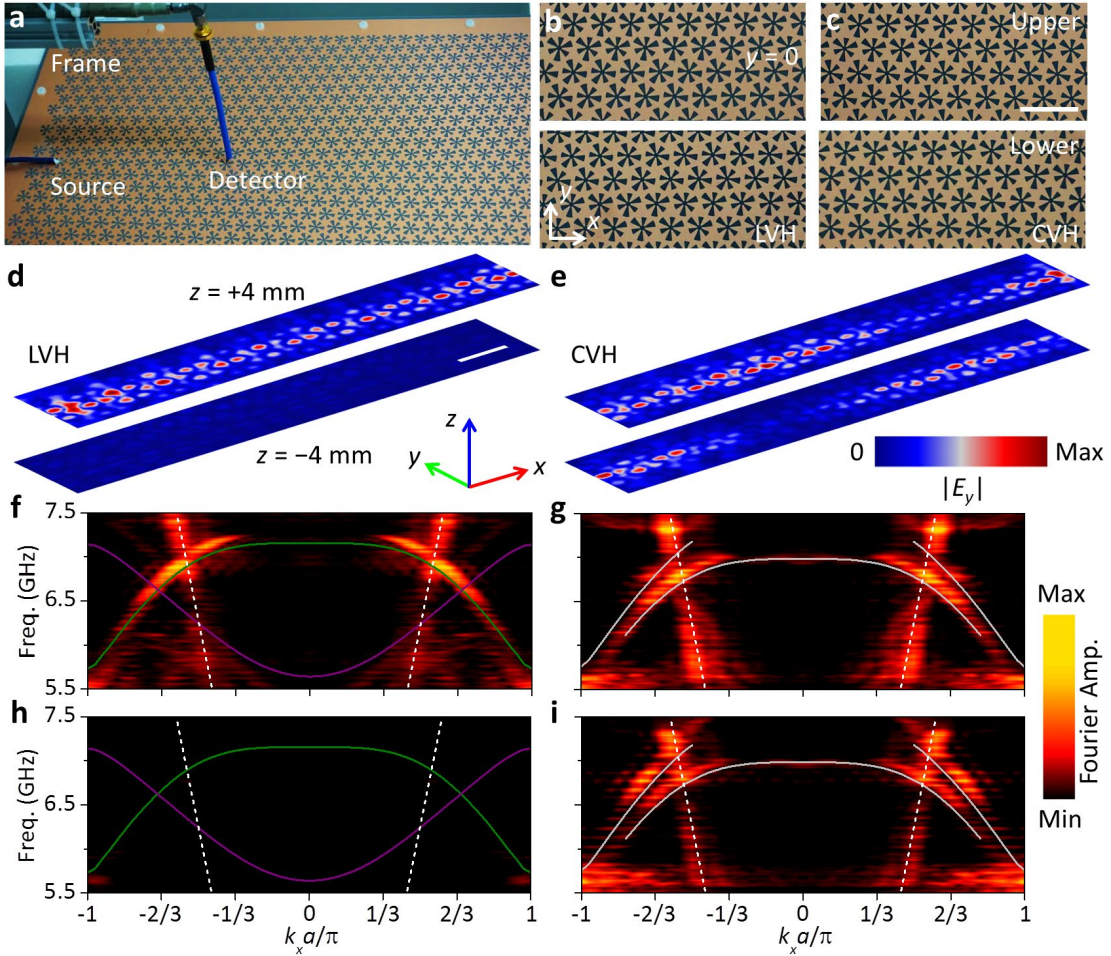

**Figure 3**. (a) Photograph of experiment setup. (b,c) Central regions of fabricated

samples. Each sample comprises two different LVH (b) or CVH (c) phases on two sides of $y$ = 0. The upper (lower) row shows upper (lower) layers, white scale bar 24 mm.

(d,e) Imaged field maps ($|E_y|$) around interfaces between LVH phases (c) or CVH

phases (d). (f-i) Layer-resolved FFT amplitude spectra, obtained from imaged fields around LVH interfaces (f,h) or CVH interfaces (g,i) on upper (f,g) and lower (h,i) slices. White dashed lines denote light cone. Colored (gray) solid lines indicate numerical dispersions of the LVH (CVH) edge states.

To directly observe these edge states, we have performed near-field mappings (see **Figure 3**(a)) with two monopole antennas used as the source and detector, respectively. The samples shown in Figure 3(b) and 3(c) are fabricated using standard printed circuit board (PCB) etching method (see Note 6 for details of the sample fabrication and experiment setup, Supporting Information). On two sides of the interface ($y = 0$) are different LVH phases ($\alpha$, $\beta$) = ($\pm 15°$, $5°$) or CVH phases ($\alpha$, $\beta$) = ($0$, $\pm 20°$). The imaged LVH and CVH edge states at 6.70 GHz are shown in Figure 3(d) and 3(e), respectively (other frequencies in Figure S5, Supporting Information). It is seen that excited LVH edge state is polarized to the upper layer where source is placed, and below a systematic study further confirms topological features of its layer pseudospins under perturbations. However, the excited CVH edge state exhibits strong fields on both slices, and the oscillation of amplitude is due to the interferences of its symmetric and anti-symmetric branches (see Figure S10, Supporting Information). We then perform fast Fourier transforms (FFT) on imaged fields and the obtained FFT amplitude spectra are shown in Figure 3(f-i). The bright regions in the layer-resolved spectra indicate dispersions of the edge states, and agree excellently with numerical calculations (solid lines). We have also measured projected bands of the bulk phases, and only uncoupled plane wave-like modes exist in the bandgap (see Figure S9, Supporting Information).

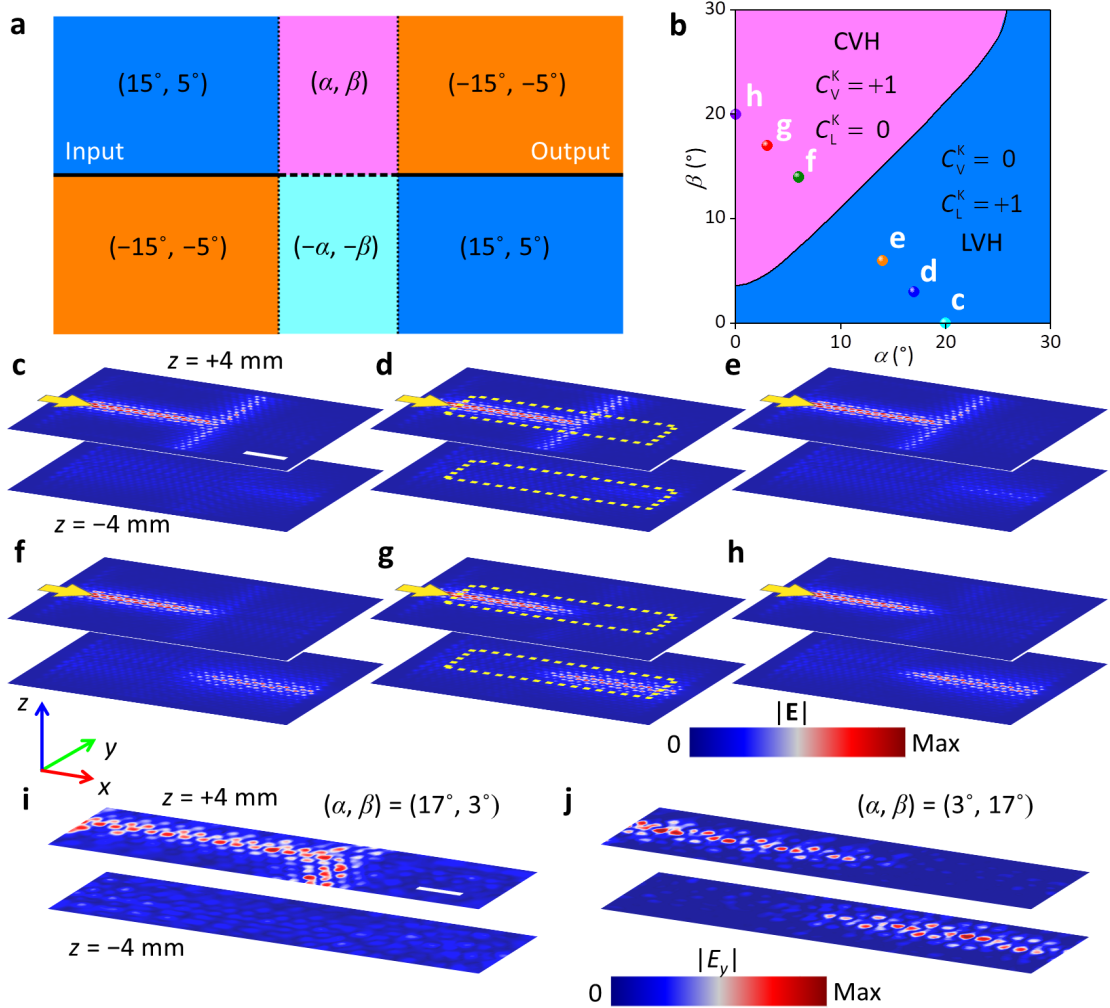

**Figure 4**. (a) Schematic diagram of the structure used to demonstrate the topological features of layer pseudospins, containing three different interfaces. The interfaces at the inlet and outlet belong to different LVH phases, and the interface in the middle is formed by the defect regions with the angles parameters $(\alpha, \beta)$ and $(-\alpha, -\beta)$ (b) Chosen values of $(\alpha, \beta)$ denoted in the phase diagram of the composite DSP crystal. (c-h) Simulated field maps with $(\alpha, \beta)$ varying to chosen values indexed by letters in (b). $(\alpha, \beta)$ in (c-e) belongs to the LVH phase, and in (f-h) belong to the CVH phase.

Yellow arrows denote the point source. The white scale bar is 60 mm. The yellow dashed rectangles denote the regions imaged in experiments. (i,j) Imaged field maps ($|E_y|$) when $(\alpha, \beta)$ = (17°, 3°) belongs to the LVH phase (i), which corresponds to simulation (d); or $(\alpha, \beta)$ = (3°, 17°) belongs to the CVH phase (j), which corresponds to simulation (g). The white scale bar is 30 mm.

Then, to demonstrate the topological protection of layer pseudospins, we consider a structure shown in **Figure 4**(a), which comprises three interfaces formed by topologically different phases. At the inlet and outlet, both interfaces are formed by opposite LVH phases, but at the inlet (left side), the forward mode at K valley is of upper layer pseudospin, while it is of lower layer pseudospin at the outlet (right side). A defect region is introduced in the middle of structure, comprises a interface formed by two phases with the angle parameters $(\alpha, \beta)$ and $(-\alpha, -\beta)$. We then smoothly perturb $(\alpha, \beta)$, with six selected $(\alpha, \beta)$ values denoted in the phase diagram (Figure 4(b)). The simulation results are summarized in Figure 4(c-h). When $(\alpha, \beta)$, the angle parameters of the defect region, belongs to the LVH phase, there is no obvious conversion of the layer polarization and the transmission to the outlet is heavily suppressed (Figure 4(c-e)). However, when $(\alpha, \beta)$ is perturbed to the CVH phases, strong fields suddenly emerge around the outlet on the lower metallic layer (Figure 4(f-h)). Such abrupt change is a key indicator of topological properties [43], and hence it confirms the topological nature of layer pseudospins. In general, the layer pseudospins are preserved if the defect does not introduce CVH phases, which lack layer-chirality and effectively mixes layer polarizations.

Further, to experimentally validate this topological protection, we fabricate two samples to experimentally confirm the results. In the first sample the defect $(\alpha, \beta) =$ (17°, 3°) belongs to the LVH phase, while in the second sample the defect $(\alpha, \beta) = (3°,$ 17°) belongs to the CVH phase. They correspond to simulated fields shown in Figure 4(d) and 4(g), and their experimentally imaged fields are shown in Figure 4(i) and 4(j), respectively. The experimental results confirm that the mixing of layer pseudospins caused by CVH phases assists the edge state in tunneling to the lower layer, while the preservation of layer pseudospins by LVH phases indeed prevents it, as we have found in numerical studies (Figure 4(c-h)).

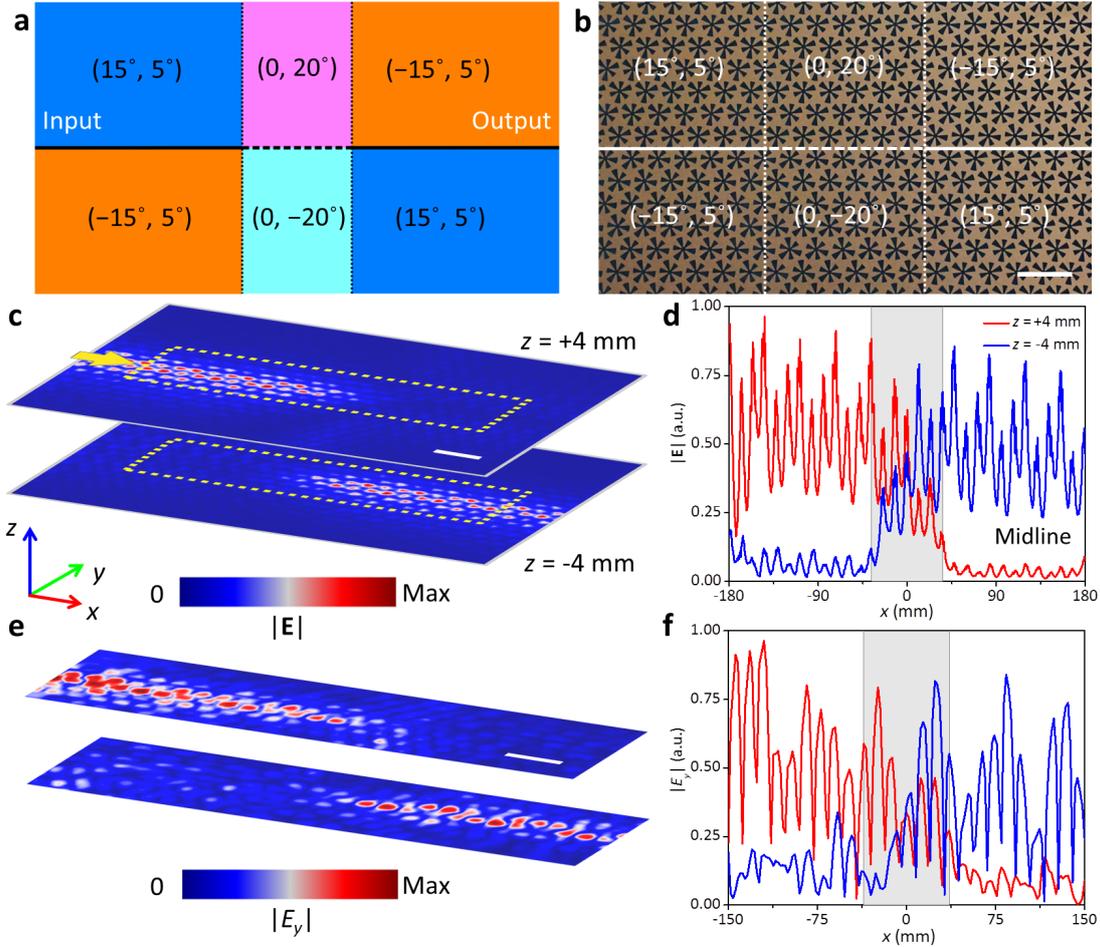

**Figure 5**. (a) Diagram of layer convertor. (b) Central region of fabricated sample. (c) Simulated field maps ($|\mathbf{E}|$) at 6.70 GHz on slices $z = \pm 4$ mm, intensities along midline ($y$=0) shown in (d), white scale bar 40 mm, shaded region denoting range of CVH phases. Yellow arrow represents the source. Yellow dashed rectangles denote regions imaged in experiments. (e) Imaged field maps ($|E_y|$) at 6.70 GHz, intensities along midline shown in (f), white scale bar 30 mm.

After the behaviors of layer pseudospins in different phases thoroughly investigated, it is time to exploit this topological DOF to design interesting photonic devices. First, we note the layer-chirality of LVH edge states and the lack of it between CVH phases. This contrast suggests a layer convertor using CVH phases as the bridge to compel an interlayer transport, and the schematic diagram is shown in **Figure 5(**a). A photograph of the fabricated sample is shown in Figure 5(b). First, we perform simulations for this device and the simulated field maps at 6.70 GHz are shown in Figure 5(c) (other frequencies shown in Figure S6, Supporting Information). To characterize the topological protection on interlayer transports, the simulated field profile along the midline is explicitly shown in Figure 5(d). Simulated temporal dynamics of the transport also show a negligible backscattering (see Figure S11 and Movie S1 for time-domain demonstration of this process, Supporting Information). We then perform near-field mappings with aforementioned experiment setup and image the region enclosed by the yellow dashed rectangle in Figure 5(c). The measured field maps at 6.70 GHz are shown in Figure 5(e) and the its field profile along the midline is shown in Figure 5(f). The shaded regions in Figure 5(d) and 5(f) denote the interval occupied by transitional CVH phases. The simulation and experimental results agree quite well and clearly show that a smooth interlayer transport happens when the layer-polarized edge state passes through CVH phases where the edge state becomes layer-mixed, validating the proposed mechanism. In fact, conventional vertical couplers used for interlayer communications in photonics need geometric optimizations to refine the performance, which focus on a single working frequency[48-50]. By contrast, our topological layer convertor functions smoothly in the bandgap (see Figure S6 and S7, Supporting Information), where the transmission efficiency (see Figure S12, Supporting Information) is generally above 80% (−1.9 dB) in experiments (working bandwidth ~0.4 GHz), and these performances may be further improved through topology optimizations[51]. Therefore, our scheme could potentially simplify fabrication process of devices with interlayer communications, relaxing fabrication tolerance, and may improve their performances in a broadband.

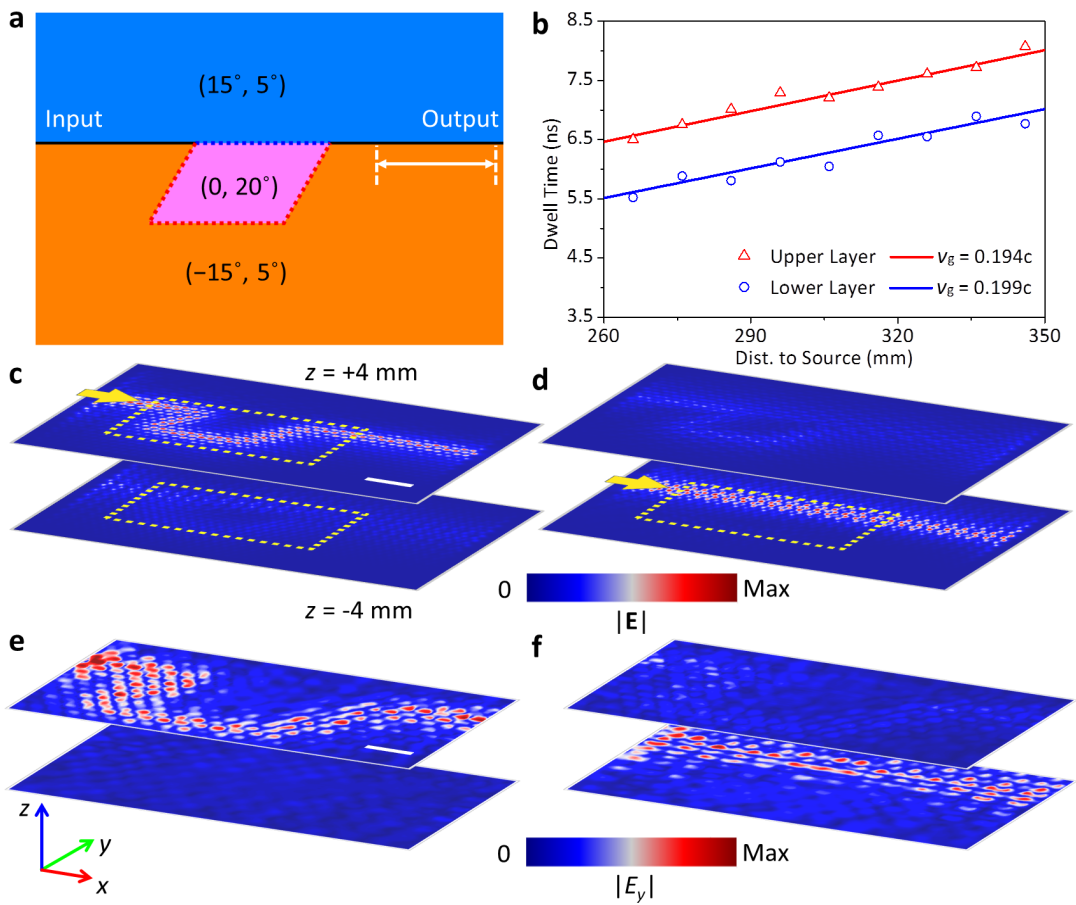

**Figure 6.** (a) Diagram of layer-selected delay line, signals excited at input and dwell time measured around output. Blue and red dashed lines denote topologically different interfaces between LVH and CVH phases. (b) Measured dwell time (scatters) when source is attached at upper or lower layer. Fitted lines manifest different intercepts, featuring the layer-selected delay. (c), (d) Simulated field maps (|**E**|) at

6.70 GHz when source (yellow arrow) is placed at upper (c) or lower (d) layer, white scale bar 60 mm. Yellow dashed rectangles denote regions imaged in experiments. (e), (f) Imaged field maps ($|E_y|$) at 6.70 GHz, white scale bar 30 mm.

Moreover, layer-polarized edge states also exist between neighboring LVH and CVH phases (see Note 4 for detailed discussion, Supporting Information). For example, when a CVH phase (0, 20°) at left meets an LVH phase (±15°, 5°) at right, the forward-going edge state at K valley is polarized to the upper (lower) layer (see Figure S4 and Table S1 for summary, Supporting Information). This intriguing interplay inspires a layer-selected topological delay line, whose schematic diagram is shown in **Figure 6**(a). A CVH phase is introduced between LVH phases as a defect, and the red (blue) dashed line highlights an interface only supporting upper (lower) layer-polarized edge states, where the forward-going one is associated with K valley (K' valley) and undergoes a (no) detour. Consequently, the excited edge state on upper (lower) layer will (not) be delayed. We then quantify this layer-selected delay by measuring their dwell time[47, 52] (see Note 6 for details, Supporting Information). The measured data plotted as functions of the distance to source (Figure 6(b)) show good linear dependence ($R^2 > 0.90$), confirming the ballistic transport of the edge states passing around the defect[52]. Further, the fitted lines exhibit similar slopes representing nearly identical group velocities, but the one on upper layer has a significantly larger intercept (~ 0.95 ns), suggesting a layer-selected delay caused by the detour. Simulated field maps for this situation (Figure 6(c) and 6(d)) show that the edge state on upper layer indeed travels through the detour while the lower one goes straightforward. We further image the regions around the detour, and experimentally obtained field maps (Figure 6(e) and 6(f)) also agree with this picture. Therefore, a layer-selected topological delay line is successfully realized and the delay on each side can be separately adjusted by tailoring shapes of the detours. Whispering-gallery-type resonators can be employed if a long dwell time is necessary[47]. Compared with previous topological delay lines[30, 52], the layer-selected delay line could offer delays in very different ranges on two layers of the device, which doubles the space utilization in a simple manner.

## 3. Conclusion

In summary, a composite DSP crystal is proposed to realize layer pseudospins in photonic valley-Hall phases. Layer-polarized valley-Hall phases emerge after deliberately twisting anisotropic slots on vertically coupled metallic layers. The layer-polarized edge states arising from these phases show fundamentally different transport behaviors compared with those found in 2D materials. Direct experimental observations, including imaged field maps and their Fourier spectra, rigorously confirm our scheme and show DSP crystals are ideal platforms to thoroughly explore the new physics. We also realize photonic devices utilizing layer pseudospins, which demonstrate more versatile control over valley-Hall edge states, in comparison with conventional devices and previous valley-Hall devices.

In future, realizing layer pseudospins in higher frequency regimes would offer new potentials for integrated photonic devices. The extension should be direct, as our scheme relies on symmetries instead of material properties and can be scaled down until the far-infrared regime, in which DSPs can still dominate over real surface plasmons[40]. For higher frequency regimes, an all-dielectric structure would be necessary to overcome thermal losses[53], but the underlying design principle for our scheme, which is based on breaking of mirror symmetries, is still applicable. Moreover, coupling layer pseudospins with other kinds of pseudospins, such as hybridized TE/TM modes[19, 26] or dipole/quadrupole modes[20, 23], would also stimulate designs for subtle and interesting topological systems which have not been found in quantum systems.

**Supporting Information**

Supporting Information is available from the Wiley Online Library or from the author.

**Acknowledgements**

X. Wu would like to thank Prof. J. H. Jiang and Dr. R.-Y. Zhang for fruitful discussions. The work is supported by an Areas of Excellence Scheme grant (AoE/P-02/12) from Research Grants Council (RGC) of Hong Kong, and grants from Natural Science Foundation of China (NSFC) (No. 11474212), Open Fund of the State

**Conflict of Interest**

The authors declare no conflict of interest.